\documentclass[11pt]{article}        
\usepackage{epsfig}
\def\G{GeV$^2$}
\def\Q{$Q^2$}
\def\v{\vspace{.1in}}
\def\P33{$\Delta(1232)$}
\def\S11{$S_{11}(1535)$}
\def\F{\mathcal{F}}
\def\K{\mathcal{K}}

\begin{document}
\begin{centering}

{\Large Baryon Form Factors at High Momentum Transfer

and Generalized Parton Distributions}

\v
Paul Stoler

\v

\it{Physics Department, Rensselaer Polytechnic Institute, Troy, NY 12180}

\end{centering}
\v

\begin{abstract} Nucleon form factors at high momentum transfer $t$ are
treated in the framework of generalized parton distributions
(GPD's). The possibility of obtaining information about
parton high transverse momentum  components by application of GPD's
to form factors is discussed. This is illustrated
by applying an {\em ad-hoc} 2-body parton wave function
to elastic nucleon form factors $F_1$ and $F_2$, the $N\to\Delta$
transition magnetic form factor $G^*_M$, and wide angle Compton scattering
(WACS) form factor $R_1$. 
\end{abstract}  

\v

\section{Introduction.}
\subsection{Valence PQCD.}
One of the most studied questions relating to
the properties of hadrons during the 
past decades has been how to describe exclusive
reactions, and in particular electromagnetic form
factors, in experimentally accessible regions of 
energy and momentum transfer. For momentum transfers
of tens of \G, corresponding to characteristic wavelengths
of less than 0.1 fm, ordinary constituent quark or
flux tube models, which are a mainstay at much lower
momentum transfers, appear to be inadequate, especially
with increasing  momentum transfer. Improved
fits have  been possible through modification of CQM's
such as the introduction of quark form factors~\cite{cardarelli}.

During the 1980's there was considerable theoretical progress
in the description of exclusive reactions at asymptotically
high momentum transfers in terms of the
fundamental current quarks, applying valence perturbative QCD (PQCD),
~\cite{efremov,brodsky,cernyak} together with SVZ~\cite{svz} sum rules.
Among the most important 
consequences of valence PQCD are the so called constituent
counting rules, which predict relatively simple dependences
of exclusive amplitudes as functions of momentum transfer.
Many reactions experimentally appear to  obey
these constituent counting rules. Also, the magnetic nucleon
elastic form factors  and some resonant transition form factors could be
roughly accounted for in magnitude through the application of
the  QCD sum rules~\cite{cernyak,carlson,stoler,sterman}, leading
to the possibility that valence PQCD would be applicable
at kinematic conditions achieved at current or planned
accelerator facilities. However, it had also been pointed
out~\cite{rad1,islls} that the results of utilizing  PQCD
sum rules led to seemingly unrealistic
valence quark longitudinal momentum fraction distributions
$\phi(x)$. When utilized 
in the PQCD calculations, these $\phi(x)$'s led to inconsistent
application  of PQCD. 

Another prediction  of PQCD is hadron 
helicity conservation. Recent experiments~\cite{frolov, perdrisat}
at  Jefferson Lab (JLab), which have measured helicity non-conserving
amplitudes for elastic and resonance  form factors, have  
shown that at momentum transfers approaching 6 \G\ the
approach to PQCD is not manifest.

\subsection{Generalized Parton Distributions.}
\begin{figure}[t]
\centerline{\epsfig{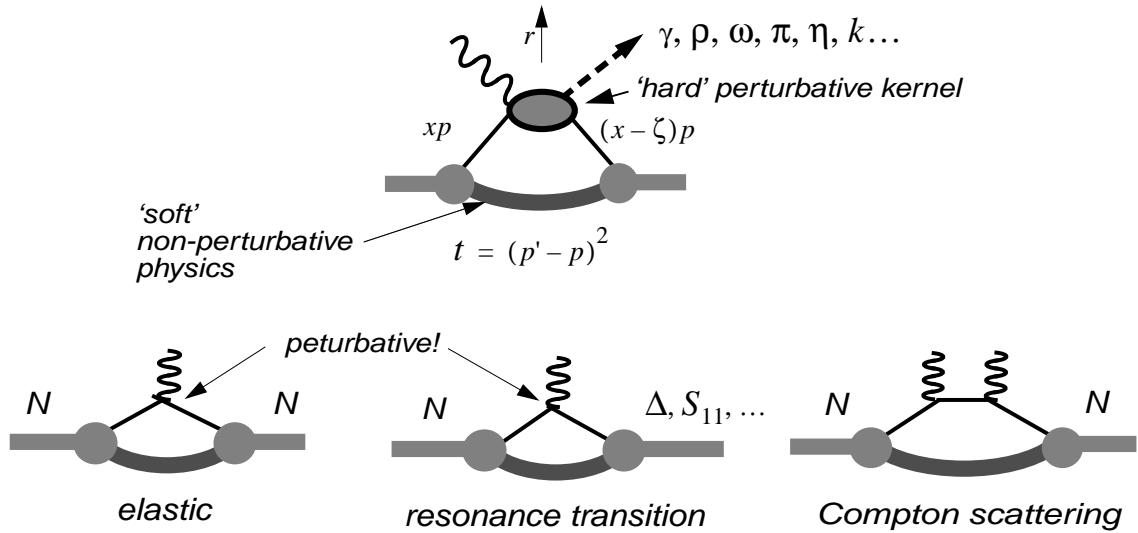}}
\caption{\label{handbag} 
Diagrams representing the ``handbag'' mechanism; upper, deeply virtual
Compton scattering and meson production; lower, nucleon elastic
and resonance form factors, and  wide angle Compton scattering.}
\end{figure}

The recent evolution of the theoretical formalism
of generalized parton distributions~\cite{ji, radgpg, collins}
(GPD's) has shown promise of
providing a framework for describing exclusive reactions
in terms of parton degrees of freedom without invoking
the internal hard mechanisms of valence PQCD. The GPD description has been
discussed primarily in the reactions involving deeply
virtual Compton scattering and meson production, at
as high \Q\ and small $t$ as possible.
This is the kinematic region in which it has been shown
that the reaction mechanism 
can be factorized into a hard perturbative, and soft non-perturbative
part~\cite{collins}, the so-called {\em handbag}
process  illustrated in fig.~\ref{handbag}.

The soft handbag is characterized
by GPD's which contain  information about the distribution
of quarks in the hadron. In particular,  they give the amplitude
that a quark with longitudinal momentum fraction $x$, in a
hadron with momentum $p$, can be given a  momentum kick
$t=r^2$ with sideways component $r_\perp$,
and re-absorbed by the hadron which emerges with a momentum
$p-r$ (see fig.~\ref{handbag}-top). Typically, hard electroproduction
processes
also require a longitudinal momentum transfer, characterized
by a skewedness parameter $\zeta \equiv r_\parallel/p$. In the limit 
$t \to 0$ the certain GPD's become identical
with the deep inelastic scattering (DIS) structure functions, while
others are not accessible in DIS.

A very important property of the GPD's are sum rules which
directly relate moments  of the GPD's
to various hadronic form factors. For example,
electron scattering form factors  are the 0'th moments of the GPD's.

For elastic scattering 
 
\begin{equation}
F_1(t)=\int^1_0\sum_q \F^q_\zeta(x,t)dx  
\label{eq:F1}
\end{equation}
\begin{equation}
F_2(t)=\int^1_0\sum_q \K^q_\zeta(x,t)dx
\label{eq:F2}
\end{equation}

\noindent where $q$ signifies both quark and anti-quark flavors.
We work in a reference frame in which the total
momentum transfer is transverse so that  $\zeta$=0, and denote
$ \F^q(x,t) \equiv \F^q_0(x,t)$, 
\ $ \K^q(x,t) \equiv \K^q_0(x,t)$, etc.

 For Compton scattering $\zeta = 0$, and 
 the appropriate {\em form factor-like}
quantities~\cite{rad_wacs} are the -1'th  moments of the GPD's

\begin{equation}
R_1(t)=\int^1_0\sum_q{1\over x} \F^q(x,t)dx 
\label{eq:R1}
\end{equation}

\begin{equation}
R_2(t)=\int^1_0\sum_q{1\over x} \K^q(x,t)dx
\label{eq:R2}
\end{equation}

Resonance transition form factors access components of the
GPD's which are not accessed in elastic scattering or WACS.
The  $N\to\Delta$ form factors are related
to isovector components of the GPD's~\cite{polyakov}. 

\begin{equation}
 G^*_M = \int^1_0\sum_q \F^q_M(x,t)dx\ \ \  
 G^*_E = \int^1_0\sum_q \F^q_E(x,t)dx\ \ \  
 G^*_C = \int^1_0\sum_q \F^q_C(x,t)dx 
\label{eq:delta}
\end{equation}

\noindent where  $G^*_M$, $G^*_E$ and  $G^*_C$ are magnetic, electric
and Coulomb transition form factors~\cite{jones}, and 
$ \F^q_M$,  $\F^q_E$, and $\F^q_C$ are axial (isovector) GPD's,
which can be related to elastic GPD's in the large $N_C$ limit
through isospin rotations. Similar relationships can be obtained 
for the $N\to S_{11}$ and other  transitions.

The GPD's as functions of
$x$  give the contributions to the form factors
due to quarks of flavor $q$ having momentum fraction $x$.
 As a function of $t$
they are directly related to the perpendicular momentum
distribution of partons in the hadron wave function
in ways which are inaccessible to DIS.  The Fourier 
transforms of the GPD's over $r_\perp$ give the transverse impact 
parameter distributions~\cite{burkardt}. 

In particular, at
high $|t|$ the resulting baryon form factors are strongly 
related to the high momentum components of  the
valence quark distribution amplitudes. This is illustrated
in the following sections, in which a simple {\em ad hoc}
power behavior of the high $k_\perp$ of the quark
distribution is used.

\section{Specific Examples.} 
\subsection{Proton Dirac Form Factor ${\bf F_1}$.}

\begin{figure}[t]
\centerline{\epsfig{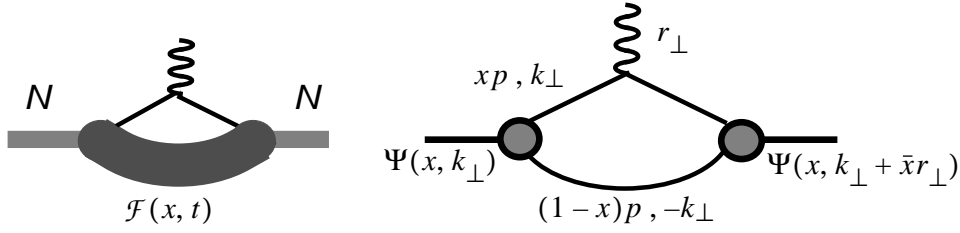}}
\caption{\label{twobody} 
Schematic representation of the handbag  mechanism as
a two-body process.}
\end{figure}

The following is based on the development by
Radyushkin~\cite{rad_wacs}, who calculated the
proton helicity conserving form factor, $F_1$, assuming
that the handbag can be expressed as an effectively
two-body process,
as illustrated in fig.~\ref{twobody}. The proton
wave function is factorized as follows

\begin{equation}
\Psi(x,k_\perp)= \Phi(x)e^{-k^2_\perp/2x\bar x \lambda^2}
\label{eq:psisoft}
\end{equation}

\noindent where $\bar x \equiv 1-x$,  and $\lambda$ is a measure of the mean 
transverse momentum.

In terms of the two-body  wave functions eq.~(\ref{eq:psisoft}), the
form factors are then expressed as:

\begin{equation}
F^{tb} = \int {\Psi^{*}(x,k_\perp + \bar x r_\perp)\Psi(x,k_\perp)
{{d^2k_\perp}\over{16\pi^3}}}
\label{eq:F1tb}
\end{equation}

\noindent Comparing eqs.~(\ref{eq:F1tb}) and (\ref{eq:F1}) gives

\begin{equation}
\F(x,t) =\int_0^1 dx  \int {\Psi^{*}(x,k_\perp + \bar x r_\perp)\Psi(x,k_\perp)
{{d^2k_\perp}\over{16\pi^3}}}
\label{eq:Ftb}
\end{equation}

\noindent Insertion of eq.~(\ref{eq:psisoft}) into eq.~(\ref{eq:Ftb}),
and evaluating the integral then gives

\begin{equation}
\F(x,t)= 
{{x\bar x \lambda^2}\over{16\pi^2}}\Phi^2(x)
e^{-\bar x t/4x \lambda^2}\equiv f(x)e^{-\bar x t/4x \lambda^2}
\label{eq:fsoft}
\end{equation}

\noindent with valence quark distributions

\begin{equation}
 f(x)= \sum_q e_q f^v_q(x)= e_uf^v_u(x)+ e_df^v_d(x) .
\label{eq:fsum}
\end{equation}

\noindent The functions $f_u(x)$ and $f_d(x)$ are chosen to agree with
the valence quark distributions $f^v_q(x)$ obtained directly in DIS. 
In particular, ref.~\cite{rad_wacs} uses an empirical function
found to agree with DIS

\begin{equation}
f^v_u(x)=1.89x^{-0.4}(1-x)^{3.5}(1+6x)
\label{eq:fvu}
\end{equation}

\begin{equation}
f^v_d(x)=0.54x^{-0.6}(1-x)^{4.2}(1+8x).
\label{eq:fvd}
\end{equation}

\noindent The function $\Phi(x)$ in eq.~(\ref{eq:phi}) is then written

\begin{equation}
\Phi_q^2(x)={{16\pi^2}\over{\lambda x\bar x}}f^v_q(x).
\label{eq:phi}
\end{equation}

\noindent Substitution of eqs.~(\ref{eq:fsum}) and ~(\ref{eq:phi})
into  eq.~(\ref{eq:fsoft}), and then  eq.~(\ref{eq:fsoft})
into  eq.~(\ref{eq:F1})
yields for the proton Dirac  form factor,
 
\begin{equation}
F_1(t)=\int^1_0\left[e_uf^v_u(x)+e_df^v_d(x)\right]
e^{-\bar x t/4x\lambda^2}dx
\label{eq:F1soft}
\end{equation}
 
\noindent The only free parameter in the analysis is $\lambda$, which 
is a measure of the mean $k_\perp$. A good fit to SLAC data for $F_1$ up to
$|t|\sim$ 8 \G\ was obtained. The resulting value of 
$\lambda^2\sim 0.7$ GeV$^2$ leads to a reasonable values
of the mean square $k_\perp$ distributions: 
$\langle k_{\perp}^2\rangle \sim (270 {\rm MeV)^2}$

\subsection{Dirac Form Factor ${\bf F_1}$ at High ${\bf t}$}

\begin{figure}[t]
\centerline{\epsfig{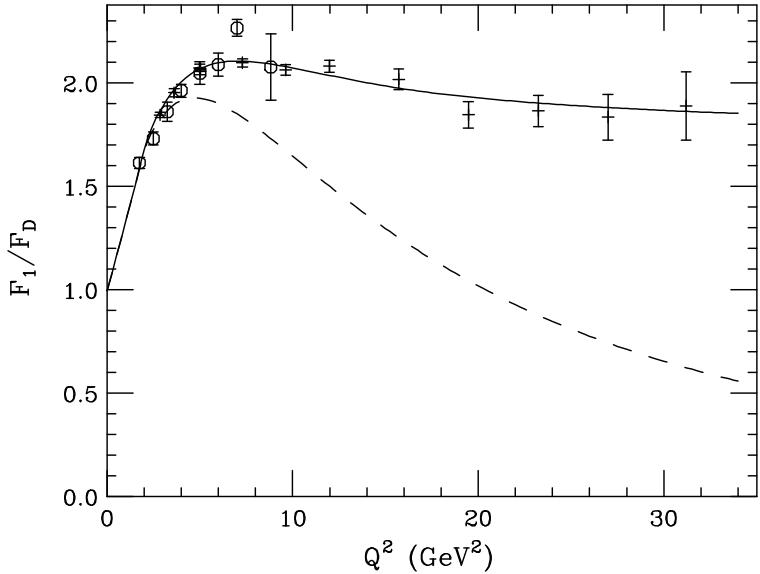}}
\caption{\label{F1} 
Proton Dirac form factor $F_1$ as a function of \Q. 
The data are from SLAC; circles ($\circ$) at lower \Q\ are
from ref.~\cite{andivahis},
and pluses (+) at higher \Q\ are  from ref.~\cite{arnold}. The dashed
curve is the result of the soft wave function $\Psi_{soft}$. The solid curve
is the result of adding a small hard component $\Psi_{hard}$ }

\end{figure}

Data on $F_1$ exists~\cite{arnold} up to a \Q\  (= $-t$) of 32 \G.
Continuation of the calculated $F_1$ to higher \Q\ exhibits a steadily
greater discrepancy with the data with increasing  \Q. This is
shown in fig.~\ref{F1}.

A subsequent ~\cite{diehl1,diehl2} elaboration of the procedure
of ref.~\cite{rad_wacs} studied the relationship between model 
quark distribution amplitudes and form factors  and Compton scattering. 
Ref.~ \cite{diehl1} used a valence quark distribution which evolves
to the  asymptotic  form \cite{brodsky,bolz},  
$\Phi(x) \to \Phi^{as}(x) \propto 120x_1x_2x_3$.
In addition to the valence $qqq$ configuration, higher Fock state 
contributions containing $qqq, g$  and $qqq,\bar q q$ were added.
For the $ k_{\perp}$ distribution,
an exponential form similar to that in ref.~\cite{rad_wacs} was used.
For simplicity, the same  $\lambda^2$ was used  for all
the three Fock state components, although the  need to consider
different  $\lambda^2$'s for the valance and higher Fock states
is discussed.
It was found that the inclusion of the higher Fock states is
significant, and can effect the form factors by as much as 20\%.
Overall the fit at higher \Q\  is improved, the fit at lower \Q\ 
somewhat deteriorated.

\begin{figure}
\centerline{\epsfig{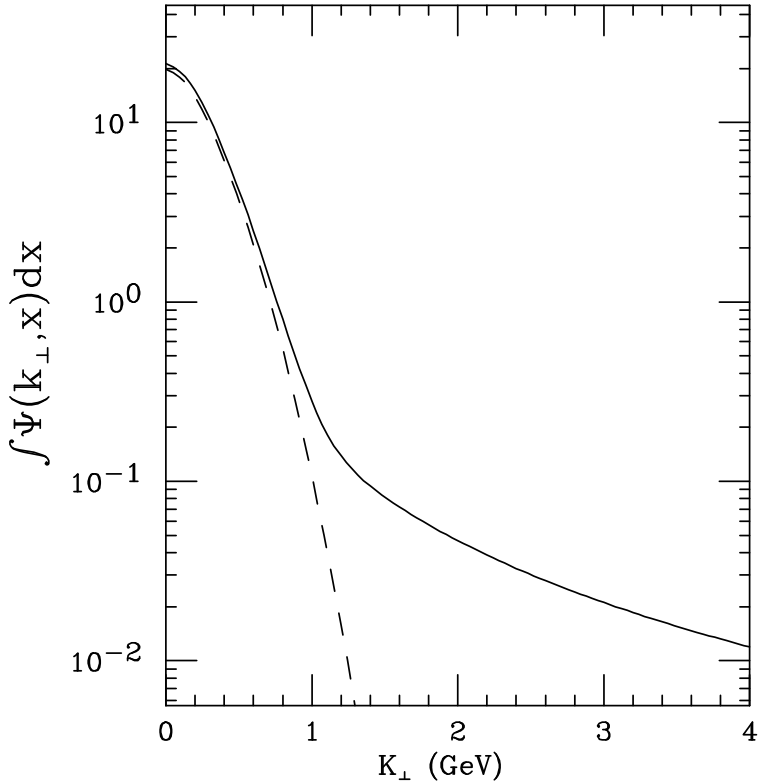}}
\caption{\label{w} 
The function $\Psi(k_\perp) \equiv \int{\Psi(x, k_\perp) dx}$ 
vs. $k_\perp$. The dashed curve
is due to the  soft Gaussian component $\Psi_{soft}$, with
$\lambda^2 = 0.7\ {\rm GeV^2}$. The solid curve is 
$\Psi_{soft} + \Psi_{hard}$, with $A_h$ = 0.065, 
$k_{\perp ,max}$ = 4 GeV, and
cutoff parameter $\Lambda$ = 0.35 GeV.}
\end{figure}

A Gaussian form of $\Psi$  cannot account simultaneously
for the  $F_1$ magnitude and shape over the entire range of \Q.
However,the addition of a small high $k_\perp$
component in eq.~(\ref{eq:psisoft}) can dramatically improve the fit.
As an example, we choose a {\em ad-hoc} $1/k_\perp^2$  behavior with
lower cutoff parameter $\Lambda$, and upper cutoff $k_{\perp,max}$:

\begin{equation} 
\Psi(x,k_\perp) = \Phi(x)\left( A_s e^{-k_\perp^2/2x\bar x \lambda^2}
+ A_h {
{x\bar x \theta(k_\perp^2<k_{\perp,max}^2)}
\over
{k_\perp^2 + \Lambda^2}
 } \right) \equiv \Psi_{soft}+\Psi_{hard}\label{eq:psihard}
\end{equation}

\noindent where $\lambda^2 = 0.7$ \G\ is fixed by
the low to intermediate \Q\ behavior of $F_1$, 
$k_{\perp,max} =4$\ \G,  and $\Lambda = 0.35$\ \G.
As seen in fig.~\ref{F1}, this small addition of $\Psi_{hard}$
in eq.~(\ref{eq:psihard}) can account for the high, as well
as the low \Q\ magnetic form factor. A value of $A_h/A_s$ = 0.065
was used in  eq.~(\ref{eq:psihard}) to obtain the solid curve. 
In all cases, the condition $\F(x,0)=f(x)$ is  maintained.
Also, note that $F_1(0) = \int\F(x,0)dx \approx 1$, 
which, is required in the definition of  $F_1$.
The function $\Psi(k_\perp)= \int{\Psi(k_\perp,x)dx}$ is shown in 
fig.~\ref{w}.

\begin{figure}
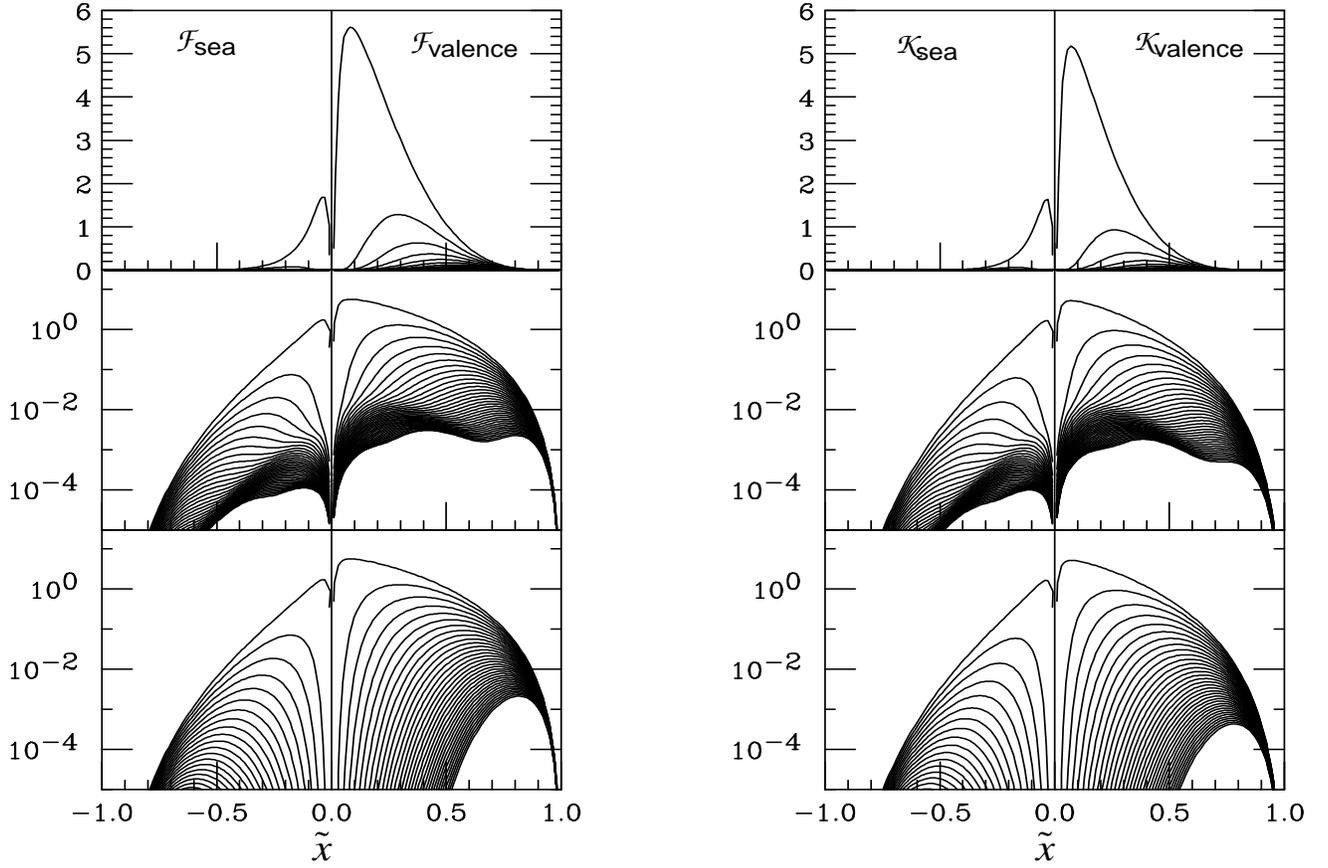

\begin{minipage}[b]{3.75in}
\epsfig{figure= F.epsi,height=4.5in,width=3.0in}
\end{minipage}
\begin{minipage}[b]{3.75in}
\epsfig{figure= K.epsi,height=4.5in,width=3.0in,angle=0}
\end{minipage}
\caption{\label{gpd} GPD's as a function of $\tilde x$ for various
values of $t$, where $\tilde x = x$ for valence quarks,
and  $\tilde x =  - x$ for the sea quarks. 
 The figures on the left and right 
are for  $\F$ and $\K$ respectively. The graphs
for  positive $\tilde x$ represent the {\em valence} quark contribution,
while the graphs for negative $\tilde x$ represent the {\em sea} quark
contributions. The full GPDs are  given by $\F = \F_{val}
- \F_{sea}$  and $\K = \K_{val} - \K_{sea}$ respectively. The
individual curves range from $|t|$  $\sim 0$ GeV$^2$ (highest curve
in each panel)   
to  $|t|$  = 35  GeV$^2$ (lowest curve in each panel). 
The upper and middle panels
are the GPD's for the full wave function $\Psi$ given in 
eq.~(\ref{eq:psihard}),
while those in the lowest panels are obtained using the
soft wave function as in eq.~(\ref{eq:psisoft}). Note that the addition
of the $\Psi_{hard}$ mainly affects the GPD's at higher $|t|$
and $\tilde x < 0.5$  }
\end{figure}

The resulting GPDs as a function of $x$ for different values
of $t$ are shown in fig.~\ref{gpd}. It would be interesting
to see how this transforms into the spacial impact distributions
$f(x,b_\perp)$ discussed in ref.~\cite{burkardt}.

\begin{figure}
\centerline{\epsfig{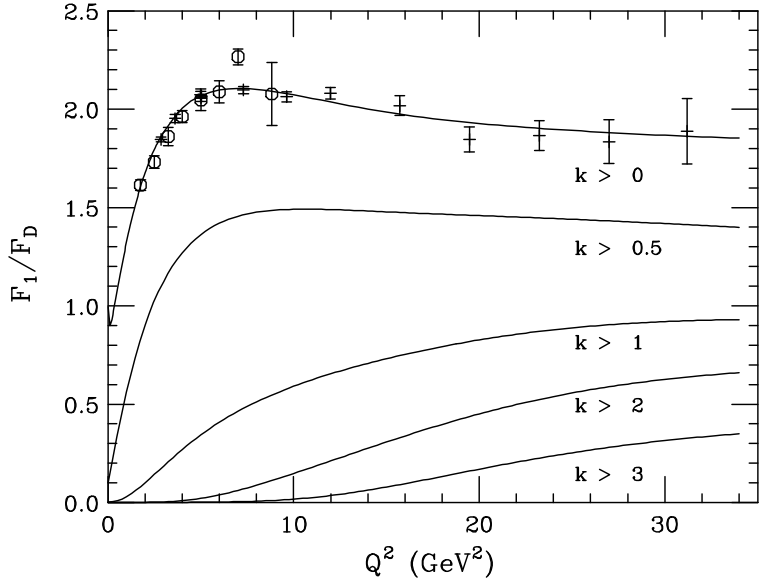}}
\caption{\label{F1k} 
The contribution to $F_1$ from different regions in $\Psi(k_\perp)$
of $k_\perp$. Each curve represents the contribution from
regions of $k_\perp$ greater than the value indicated.
The data are from SLAC; circles ($\circ$) at lower \Q\ are
from ref.~\cite{andivahis},
and pluses (+) at higher \Q\ are  from ref.~\cite{arnold}.}
\end{figure}

Figure~\ref{F1k} shows the contribution to $F_1$ from different
$k_\perp$ regions. It is clear that the
high \Q\  regions of $F_1$ are selective of the high components
of $\Psi(k_\perp)$.

\subsection{Pauli form factor ${\bf F_2}$}

Applying the analogous formalism as in ref.~\cite{rad_wacs},
Afanasev~\cite{afan} modeled $\K(x,t)$ in eq.~\ref{eq:F2}
as 

\begin{equation}
\K^q(x,t)=k^q(x)e^{-\bar x t/4x \lambda^2}.
\label{eq:Ktb}
\end{equation}

\noindent with $\lambda$ and the normalization at \Q\ = 0
free parameters.

Unlike the case for $f^q(x)$, an  expression for $k^q(x)$
cannot be obtained from DIS.
Reference~\cite{afan} notes that asymptotically PQCD and
the SVZ sum rules require an extra factor of $1-x$ for $k^q(x)$.
In the present study, it is found that  the simplest form
$k^q(x)=(1-x)f^q(x)$ adequately describes the data.
The expression for $F_2$ is then
 
\begin{equation}
F_2(t)=\int^1_0\left[e_uk^v_u(x)+e_dk^v_d(x)\right]
e^{-\bar x t/4x\lambda^2}dx
\label{eq:F2soft}
\end{equation}
 
\begin{figure}
\centerline{\epsfig{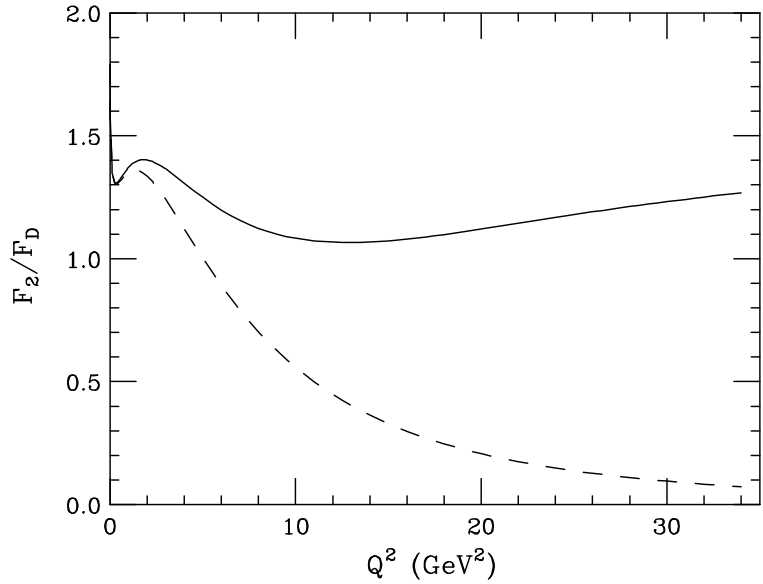}}
\caption{\label{F2} 
Proton Pauli form factor $F_2$ as a function of \Q . The dashed
curve is the result of inclusion only of 
$\Psi_{soft}$ in eq.~(\ref{eq:psihard}).
The solid curve is the result of adding  $\Psi_{hard}$ }
\end{figure}

Figure~\ref{F2} shows $F_2$, and fig.~\ref{Q2F2F1}
shows $Q^2F_2(Q^2)/F_2(Q^2)$ compared 
with the recent JLab data~\cite{perdrisat}.
 The obtained $G_{EP}/G_{MP}$, also compared
with the recent JLab data is shown in fig.~\ref{gegm}. For $F_2$
the best fit was obtained for $\lambda^2$ = 0.5, and $F_2(0)$
was normalized to $\kappa=1.79$. All other parameters,
including $A_H$ and $\Lambda$ were fixed by $F_1$.
The contribution 
of $\Psi_{soft}$ as in eq.~(\ref{eq:psisoft})
is also shown. The effects of inclusion of 
$\Psi_{hard}$ become important at around \Q\ = 8 GeV$^2$.
The JLab upgrade includes measurements to about 15 GeV$^2$, which
should test this.

\begin{figure}
\centerline{\epsfig{figure=Q2F2F1.epsi, height=10cm, angle=90}}
\caption{\label{Q2F2F1} 
The ratio of the proton Dirac and Pauli form factors $Q^2F_2/F_1$
as a function of \Q . The dashed
curve is the result of including only $\Psi_{soft}$. The solid curve
is the result of adding $\Psi_{hard}$.  The data are the recent
JLab results~\cite{perdrisat}.}
\end{figure}

\begin{figure}
\centerline{\epsfig{figure=gegm.epsi, height=10cm, angle=90}}
\caption{\label{gegm} 
The ratio of the proton electric and magnetic
form factors $G_E/G_M$  as a function of \Q. The dashed
curve is the result of  inclusion only of $\Psi_{soft}$. The solid curve
is the result of adding $\Psi_{hard}$. The data are the recent
JLab results~\cite{perdrisat} }
\end{figure}

\subsection{Wide angle Compton Scattering}

The inclusion of $\Psi_{hard}$ in eq.~(\ref{eq:psihard})
also directly affects wide angle Compton scattering (WACS).
Analogous to the elastic
electron scattering form factors $F_1$ and $F_2$ are the 
Compton ``form factors'' $R_1$ and $R_2$, which are related
to the Klein-Nishina cross section:

$$ {{d\sigma}\over{dt}} =\left({{d\sigma}\over{dt}}\right) _{KN}R^2$$

\noindent with $R^2 = R_1^2(t) + {{-t}\over{4m_p^2}}R_2^2(t)$.
As in the case with electron scattering,  
$R_2(t)/ R_1(t)$ is expected to fall as a power of $|t|$, so that
the cross section at high $|t|$ is dominated by $R_1(t)$.

Since the integrals in $R_1$ and $R_2$ (eqs.~(\ref{eq:R1},~\ref{eq:R2}) are
weighted with $1/x$, they may be expected to be more sensitive
to the sea quark distribution. Reference~\cite{rad_wacs}
adds a sea quark contribution to $f_{Compton}$:

$$f_{Compton} = f_{Compton}^{val} + f_{Compton}^{sea}$$

\noindent with

$$ f_{Compton}^{val}=e_u^2f^v_u(x)+e_d^2f^v_d(x)$$

\noindent and

$$ f_{Compton}^{sea} =(e_u^2 +e_d^2 +e_s^2)f^{sea}(x)$$

\noindent where $f^v_u$ and $f^v_d$ are parameterized
as in eqs.~(\ref{eq:fvu}) and (~\ref{eq:fvd}), respectively, and
$f^{sea}(x)$ is parameterized as follows:

$$f^{sea}(x) =0.5x^{-0.75}(1.0-x)^7$$

\begin{figure}
\centerline{\epsfig{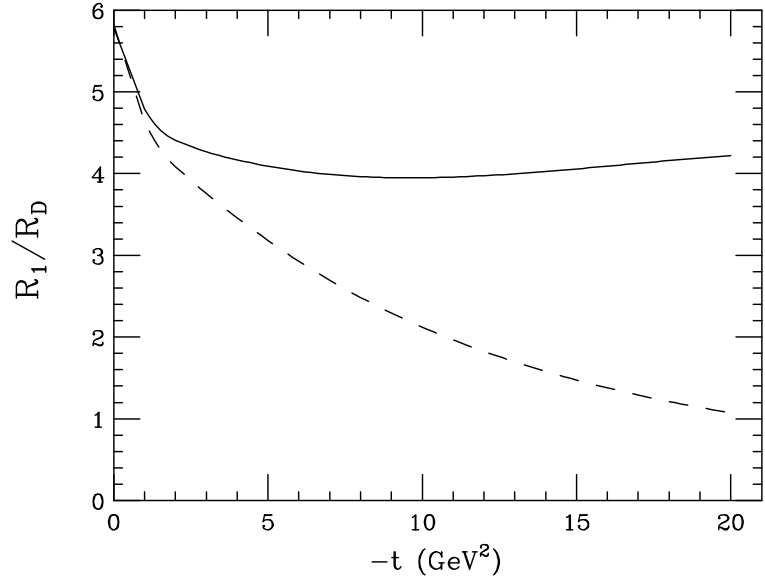}}
\caption{\label{rsum} 
The Compton form factor $R_1(t)$ vs. $t$,
using $\Psi=\Psi_{soft}$ (dashed) and 
 $\Psi=\Psi_{soft}+\Psi_{hard}$ (solid). }
\end{figure}

\begin{figure}
\centerline{\epsfig{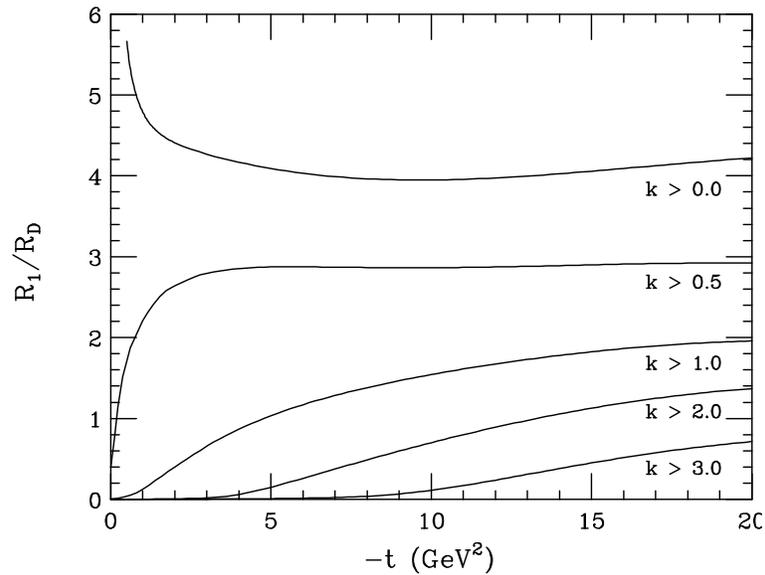}}
\caption{\label{R1k} 
The contribution to $R_1$ from different regions in $\Psi(k_\perp)$
of $k_\perp$. Each curve represents the contribution from
regions of $k_\perp$ greater than the value indicated.}
\end{figure}

Figure~\ref{rsum} shows the result for  $R_1(t)$ 
with and without
the presence of  $\Psi_{hard}$. All parameters in $\Psi_{soft}$
and $\Psi_{hard}$ are fixed by the fit to $F_1(Q^2)$.
The proposed experiments at JLab with a 12 GeV electron
beam are expected to reach $|t|$ = 15 \G, and therefore should
be sensitive to the consistency of the approach.
Figure~\ref{R1k} shows the contribution to $R_1$ from different
$k_\perp$ regions of $\Psi(k_\perp)$. It is clear that the
high $|t|$ regions of $R_1$ are selective of the high components
of $\Psi(k_\perp)$.

Reference~\cite{diehl1} have also evaluated wide angle Compton 
scattering, and find the
higher Fock components contribute an even more significant
fraction of the form factors ($R_1,R_2$) than in elastic scattering,
especially at lower $|t|$, as may be expected.

\subsection{$N\to \Delta$ transition.}

The transition $N\to \Delta(1232)$ is purely isovector, which
can be expressed in terms of three transition from 
factors~\cite{jones}; 
magnetic $G_M^*(Q^2)$, electric  $G_E^*(Q^2)$, and 
Coulomb (or scalar) $G_C^*(Q^2)$, with $(p_\Delta-p)^2 = t = -Q^2$.
These can be expressed in terms of the isovector components of
the GPD's~\cite{polyakov}:



$$ G_M^*(t)=\int^1_0\sum_q \F_M^{(3)q}(x,t)dx$$
$$ G_E^*(t)=\int^1_0\sum_q \F_E^{(3)q}(x,t)dx$$
$$ G_C^*(t)=\int^1_0\sum_q \F_C^{(3)q}(x,t)dx$$

\begin{figure}
\centerline{\epsfig{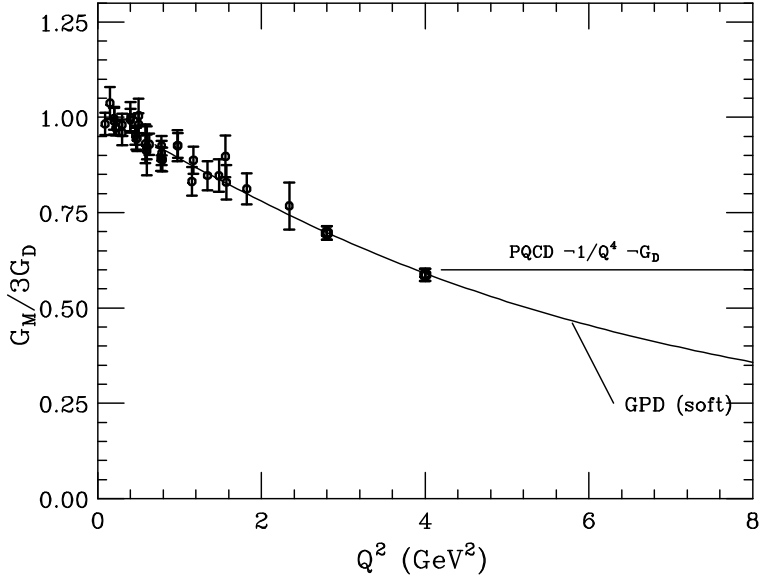}}
\caption{\label{gmdelta} 
The $N\to\Delta$ magnetic form factor  $G_M^*(Q^2)$ relative
to the dipole  $G_D=3/(1+0.71Q^2)^2$. The data points for
\Q\ below 2.8 \G\ are from a compilation of ref.~\cite{kamalov}.
Those at \Q\ = 2.8 and 4.0 \G\ are recent JLab data~\cite{frolov}.
The horizontal line reflects the $1/Q^4$ asymptotic PQCD shape,
and the curve  denoted GPD is  discussed in the text.}
\end{figure}

The experimental status of the $N\to\Delta$ transition 
magnetic form factor is shown in figure~\ref{gmdelta}.

\noindent The decrease in  $G_M^*$ relative to  $G_D$ indicates that
the transition form factor is softer than that for elastic scattering.
The soft part of the
transition form factor can be modeled
by assuming a different $\lambda$ parameter for $\Psi_\Delta$ than  
$\Psi_P$. Thus, we take

$$\Psi_{P,soft}(x,k_\perp)= \Phi_P(x)e^{(-k^2_\perp/2x\bar x \lambda_P^2)}$$

$$\Psi_{\Delta,soft}(x,k_\perp)= 
\Phi_\Delta (x)e^{(-k^2_\perp/2x\bar x \lambda_\Delta^2)}$$

\noindent For simplicity we have taken
$\Phi_\Delta(x)=\Phi_P(x)=\Phi(x)$. This leads to 

$$F_{N\Delta}(x,k_\perp,t)=
{{x\bar x}\over{8\pi^2}}
\left({{\lambda_P^2\lambda_\Delta^2}\over {\lambda_P^2+\lambda_\Delta^2}}\right)\Phi^2(x)
e^{-{{\bar x |t|}\over{2x}}
\left({{1}\over {\lambda_P^2+\lambda_\Delta^2}}\right)
.}$$

The curve passing through the data is obtained with 
$\lambda^2_{P\Delta}\equiv \lambda_P^2+\lambda_\Delta^2 = 0.38$ 
implying a transition $(k_\perp)_{RMS} \sim (180)^2\ {\rm (MeV)^2} $, 
which is
considerably smaller than the value obtained for elastic scattering
from the  proton. This also implies that the mean transition
radius for the $N\to\Delta$ transition  is also larger than for proton elastic
scattering.

\section{Conclusion.}

The advent of the GPD formalism offers a framework to
model the $k_\perp$ distributions of quarks which
are involved in exclusive reactions. These
distributions are constrained by providing
simultaneous fits to several different reactions rather than
by fitting form  factors for a single specific reaction.
Furthermore, specific reactions may be sensitive to
specific components of GPD. For example the $N\to \Delta$ is
selective of isovector components.

Within the two-body framework presented, high \Q\ (or$|t|$)
form factors are  sensitive to the high momentum components of the  
underlying wave functions. The sensitivities become significant
at \Q\ or $|t|$ greater than about 7 or 8 GeV$^2$. Currently, only
 $G_{MP}$ experimental data extend to much higher values
of \Q. However, the proposed program of high $|t|$ exclusive
measurements for the Jefferson Lab 12 GeV upgrade is
anticipated to provide high quality data 
for all of the reactions discussed here.
 
The example given here for modeling
the {\em hard} part of the wave function is purely {\em ad-hoc}
and not meant as a rigorous theoretical procedure, but
indicative of the sensitivity of high $|t|$
exclusive reactions to high momentum components of the
nucleon parton distribution. More rigorous
theoretical approaches are expected in parallel with the high quality
data expected in the future.

\v
  
\noindent{\em Acknowledgements:}

The author thanks Anatoly Radyushkin 
for much  discussion and guidance. Richard Davidson is thanked
for helpful discussions.

 The work was partially supported by the
{\em National Science Foundation}.


\begin{thebibliography}{16}

\bibitem{cardarelli}F. Cardarelli et al., {\em Physics Letters} {\bf B371}, 7 (1996)


\bibitem{efremov}A.V. Efremov and A.V. Radyushkin, {\em Theor. Math. Phys.}
 {\bf 42}, 97 (1980)  

\bibitem{brodsky} S. J. Brodsky and G. P. Lepage, {\em Phys. Rev.} {\bf D22}, 
2157 (1980).


\bibitem{cernyak} V.L. Chernyak and I.R. Zhitnitsky, {\em Physics Reports},
{\bf 112}, 173 (1984).

\bibitem{svz}M.A. Shiffman, A.I. Vainstein and V.I.Zakharov, 
{\em Nuc. Phys.}, {\bf B47},385,448,519 (1979).

\bibitem{carlson}C. E. Carlson and J. L. Poor, {\em Phys. Rev.} {\bf D38}, 2758 (1988).

\bibitem{stoler} P. Stoler, {\em Physics Reports}{\bf 226}, 103 (1993).

\bibitem{sterman}G. Sterman and P. Stoler, Annual Reviews of 
        Nuclear and Particle Science, {\bf 47}, 193 (1997).

\bibitem{rad1} A.~V.~Radyushkin, {\em Nucl.~Phys.}~{\bf A527}, 153c (1991).

\bibitem{islls} N.~Isgur and C.~H.~Lewellyn- Smith,
         {\em Phys.~Rev.~Lett.}~{\bf 52}, 1080 (1984);

\bibitem{frolov} V. V. Frolov, PhD thesis, Rensselaer Polytechnic Institute;
V. V. Frolov, {\it et al.} {\em Phys. Rev. Lett.} {\bf 82},45 (1999).

\bibitem{perdrisat}M.K. Jones {\it et al.} {\em Phys. Rev. Lett.} {\bf 84},1398 (2000).

\bibitem{ji}X. Ji, {\em Phys. Rev. Lett.} {\bf 78}, 610 (1997); 

\bibitem{radgpg}A. Radyushkin, {\em Phys. Lett.} {\bf B380},417 (1996);
{\em Phys. Rev.} {\bf D56},5524 (1997).

\bibitem{collins}J. Collins, L. Frankfort, and M. Strikman, {\em Phys. Rev.},
{\bf D56}, 2982 (1997).

\bibitem{rad_wacs}A. Radyushkin, {\em Phys. Rev.} {\bf D58},114008 (1998).

\bibitem{polyakov}K.Goeke, M.V. Polyakov, and M. Vanderhaeghen,
Prepriint hep-ph/01060112, 1 June, (2001)

\bibitem{jones} H.F. Jones and M.D. Scadron, {\em Annals of Physics} {\bf 81}, 1 (1979).

\bibitem{burkardt} {\em Phys. Rev.} {\bf D62},0701503 (2000); hep-ph/0005108.



\bibitem{arnold} R.G. Arnold et al., {\em Phys. Rev. Lett.} {\bf 57}, 174 (1986).

\bibitem{andivahis}L. Andihavis {\em et al.}, {\em Phys. Rev.} {\bf D50}, 5491 (1994).



\bibitem{diehl1} M.Diehl, Th. Feldmann, R. Jakob and P. Kroll,
{\em Eur. Phys.} {\bf C8}, 409 (1999); hep-ph/9811253.

\bibitem{diehl2} M.Diehl, Th. Feldmann, R. Jakob and P. Kroll,
{\em Nucl. Phys.} {\bf B596}, 33 (2001), Erratum-ibid. {\bf B605}, 647 (2001);
hep-ph/0009255.


\bibitem{bolz} J. Bolz and P. Kroll,{ \em Z. Phys.} {\bf A356},327 (1996).

\bibitem{afan}A. Afanasev, E-print: hep-ph/9910565;
 ``Proceeding of the JLAB-INT Workshop on Exclusive and
  Semi-Exclusive Processes at High Momentum Transfer'', 
  C. Carson and A. Radyushkin, eds. World Scientific (2000).
  May 1999

\bibitem{kamalov}S.S. Kamalov and S. N. Yang, {\em Phys. Rev. Lett.}
{\bf 83}, 4494 (1999).

\end{thebibliography}
\end{document}